# Photonic quasicrystal of spin angular momentum

Min Lin[1], Xinxin Gou[1], Zhenwei Xie[1], Aiping Yang[2], Luping Du[1,*], Xiaocong Yuan[1]

[1]*Institute of Microscale Optoelectronics & State Key Laboratory of Radio Frequency Heterogeneous Integration, Shenzhen University, Shenzhen 518060, China*

[2]*Research Institute of Interdisciplinary Science & School of Materials Science and Engineering, Dongguan University of Technology, Dongguan 523808, China*

*Corresponding authors: Email: lpdu@szu.edu.cn.

**Quasicrystals, characterized by long-range order without translational symmetry, have catalyzed transformative advances in various fields, including optics in terms of field quasicrystals. Here, we present the first demonstration of photonic quasicrystals formed by spin angular momentum, unveiling novel spin–orbit coupling effects absent in traditional field quasicrystals. A de Bruijn tiling like theoretical framework was built elucidating the formation mechanism of spin quasicrystals for diverse symmetries. Moreover, the configurations of these spin textures can be manipulated through the adjustments of the wavefronts, among which phason-like discontinuous dynamics is observed and quantitatively measured. Unlike optical quasicrystals shaped by electromagnetic fields, these spin-governed quasicrystals exhibit quasi-periodic properties of kinematic parameters, extending their potential applications to other physical systems. These findings hold promise for novel advancements in optical trapping, quasicrystal fabrication, and optical encryption systems.**

The discovery of quasicrystals, possessing long-range order without periodicity, has ignited a revolution in crystallography. Initially studied in metallic alloys[1], quasi-periodic structures have since garnered extensive attention across various physical systems, including magnetic materials[2–5], superconductors[6], and twisted bilayer graphene[7,8]. The identification of quasicrystals in condensed-matter physics has also initiated a novel area of investigation in photonics. The influence of distinct symmetries on wave propagation in quasi-periodic structures has given rise to numerous peculiar properties, such as disorder-enhanced transport[9], localization of linear excitations[10], and the dynamics of phasons[11,12]. The higher symmetries of photonic quasicrystals, not found in their periodic counterparts, have facilitated abundant optical functionalities, including a complete photonic band gap[13], enhanced light emission[14], and imaging by negative refraction[15]. Moreover, quasi-periodic photonic systems have provided new experimental platforms for studies of low-threshold lasing[16], nonlinear frequency conversion[17], Bose-Einstein condensates[18] and optical trapping of ultracold atoms[19].

However, previous studies have primarily focused on the electric field patterns of quasi-periodic photonic systems, leaving quasi-periodic photonic spin textures largely unexplored. Recently, spin–orbit coupling in focused beams or evanescent waves has revealed a plethora of fascinating deep-subwavelength phenomena[20–22], and a variety of novel topological quasi-particles including skyrmions[23,24], merons[25], skyrmion lattices[26–29], meron lattices[28–30], hopfions[31] and topological spin defects[32] have been discovered in these photonic spin textures, which are also perceived as next-generation information carriers[33,34]. In this context, investigating various symmetries in these spin textures beyond periodic conditions is crucial for broadening the horizons of topological photonics and enhancing the control over topological configurations.

In this work, the quasi-periodic structures were integrated into the photonic spin textures for the first time, resulting in the emergence of various unusual effects attributed to the spin–orbit coupling that are absent in traditional field quasicrystals. The formation mechanism of the quasi-periodic photonic spin textures was elucidated within a theoretical framework in compliance with the generalized multigrid method, which was presented by the mathematician N. G. de Bruijn and offered a reliable algorithm for generating the quasi-periodic tiling with various symmetries[35]. This sheds light on a comprehensive understanding of quasi-periodic photonic systems with diverse symmetries, addressing a gap in previous studies that predominantly focused on photonic quasicrystals possessing specific symmetries. Furthermore, the configurations of the quasi-periodic photonic spin textures were manipulated by adjusting the wavefronts of the photonic systems, analogous to manipulating the quasi-periodic tiling configurations through adjustments of parallel line offsets in the generalized multigrid method. Within these configuration manipulations, phason-like discontinuous transports[19] were observed in quasi-periodic photonic systems, for which the displacements were quantitatively determined. The results of this work not only provide insights into the spin–orbit coupling with diverse symmetries under quasi-periodic conditions and the dynamics of quasi-periodic photonic spin textures, but also have potential applications in the field of optical traps[36], quasicrystal fabrications through optical induction[37], and the optical encryption systems[38].

The generation of the quasi-periodic photonic spin textures was realized on the interference of the transverse magnetic (TM) evanescent waves. In this photonic system, the longitudinal component of the electric field is $E_z=\sum_{n=1}^{N} A e^{ik_r\left(x\cos\frac{2n\pi}{N}+y\sin\frac{2n\pi}{N}\right)}e^{-k_z z}$, with transverse components $E_x$ and $E_y$ satisfying $E_x=\frac{1}{k_r^2}\frac{\partial^2 E_z}{\partial x \partial z}$ and $E_y=\frac{1}{k_r^2}\frac{\partial^2 E_z}{\partial y \partial z}$ [26], where $A$ is a constant, $N$ is the number of evanescent waves, $k_r$ and $ik_z$ are the transverse and longitudinal wave-vector components. The spin angular momentum (SAM) is defined as $\boldsymbol{S}=\text{Im}[\varepsilon \boldsymbol{E}^*\times\boldsymbol{E}+\mu\boldsymbol{H}^*\times\boldsymbol{H}]/4\omega$ [39], where * indicates complex conjugation, $\omega$ denotes the angular frequency of the electromagnetic field, $\varepsilon$ and $\mu$ denote the permittivity and permeability of the medium, respectively. The longitudinal component of $\boldsymbol{S}$ can be mathematically represented as $S_z=\frac{\varepsilon}{2\omega}\sum s_z'(i,j)$ (see Supplementary Section 1), which involves a superposition of standing waves, with each standing wave being calculated as:

$$s_z' = \sin\left(\theta_j - \theta_i\right)\sin\left[-k_\alpha \sin\left(\frac{\theta_i + \theta_j}{2}\right)x + k_\alpha \cos\left(\frac{\theta_i + \theta_j}{2}\right)y\right], \quad (1)$$

where $\theta_i=2i\pi/N$ and $\theta_j=2j\pi/N$ are the in-plane propagation angles of the $i^{th}$ and $j^{th}$ ($j>i$) evanescent wave respectively, $\alpha=j-i$, and $k_\alpha=2\sin(\alpha\pi/N)k_r$ is the wave-vector of the standing wave. For example, the formation of the photonic spin quasicrystal with $N=5$ is illustrated in Fig. 1a, where the interactions between each pair of evanescent waves are denoted by numerical labels. Each standing wave is formed by the interaction between the $i^{th}$ and $j^{th}$ evanescent waves, which manifests as two points positioned opposite each other in the Fourier domain and labelled with the corresponding numbers. In the case of interference of evanescent waves with $N=5$, there exist only two types of interactions: between the nearest and next-nearest evanescent waves, corresponding to $\alpha=1$ and 2 respectively. Consequently, there are two sets of wave-vectors of the standing waves, $k_1$ and $k_2$. The formation mechanism of the photonic spin quasicrystal complies with the generation of quasi-periodic tiling based on the generalized multigrid approach[35]. In the generalized multigrid method, the plane is divided into $N$ sets of parallel lines that are evenly spaced, with each set oriented at an angle of $2n\pi/N$ ($n=1, 2, \ldots, N$). Degenerate points, where more than two lines intersect, are avoided by suitably choosing offsets of the lines in the direction normal to the lines. At each intersection between two lines, a rhombus is constructed with faces normal to those parallel lines. For example, a typical region of the parallel lines with $N$=5 is illustrated in Fig. 1b. Note that there exist only two types of intersections, corresponding to thin and wide rhombus characterized by acute angles of $\pi/5$ and $2\pi/5$, respectively. Once the rhombuses are assembled, a quasi-periodic tiling is generated, which manifests as a Penrose tiling[40] in the case of $N$=5.

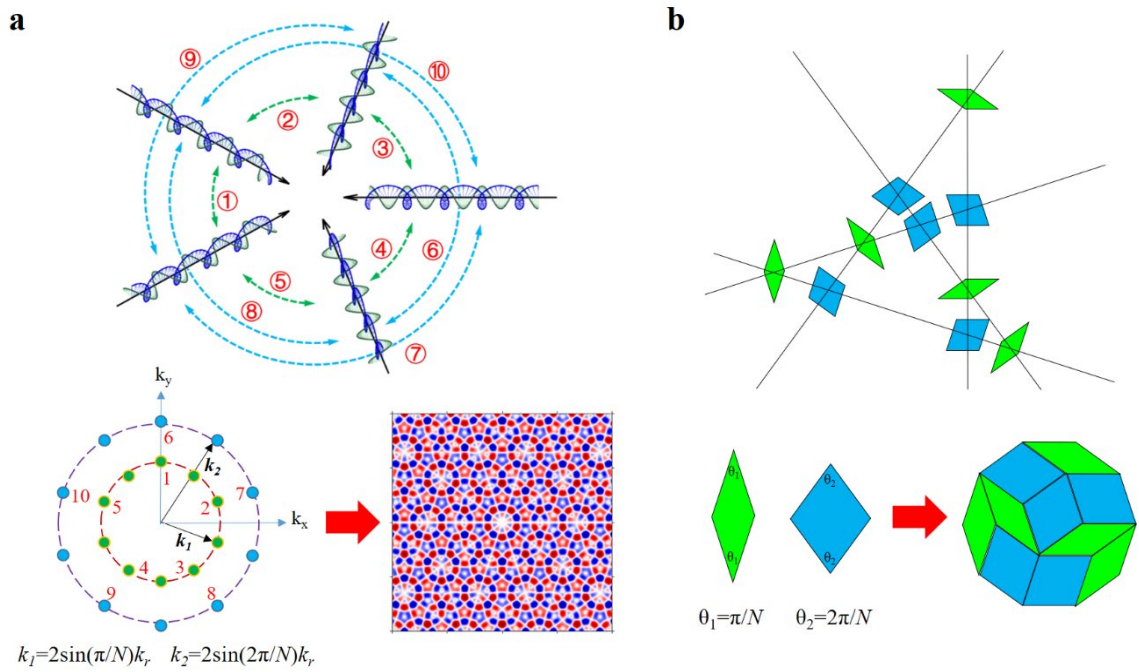

**Fig. 1 | Formation mechanism of the photonic spin quasicrystal and its correspondence to the generation of the de Bruijn tiling. a,** Schematics of the interference of the evanescent waves (top), and the corresponding $S_z$ with $N$=5 in the Fourier and the spatial domain (bottom). The correlations between each standing wave and the interaction of the evanescent waves are denoted by the numerical labels. **b,** Typical region of the parallel lines ($N$=5) with rhombuses constructed at each intersection (top), and the formation of de Bruijn tiling through assembling the rhombuses with different acute angles (bottom).

The correspondence between the generation of photonic spin quasicrystals and de Bruijn tiling is enumerated in Table 1, under the condition that $N$ is an odd integer. Note that the interaction between evanescent waves resembles the intersection between parallel lines in the generalized multigrid method, where the number of types in both cases is equal to $(N-1)/2$. Consequently, the wave-vector of the standing waves can be compared to the rhombus in de Bruijn tiling, with the number of sets and the magnitude of the wave-vector equal to the types of rhombuses and the sine of the acute angle of the rhombus, respectively, which are $(N-1)/2$ and $\sin(\alpha\pi/N)$.

**Table 1 | Correspondence between photonic spin quasicrystal and de Bruijn tiling**

| Photonic spin quasicrystal | De Bruijn Tiling | |
|---|---|---|
| Number of evanescent waves | Number of sets of parallel lines | $N$ |
| Types of interaction between the evanescent waves | Types of intersection between the parallel lines | $(N-1)/2$ |
| Sets of wave-vector of the spin textures | Types of rhombuses | $(N-1)/2$ |
| Magnitude of the wave-vector of the spin textures ($k_a/2k_r$) | Sine of the acute angle of the rhombus | $\sin(\alpha\pi/N)$ |

The aforementioned model allows for the prediction of the number of sets and the magnitude of the wave-vector, thereby enabling the calculation of the number and sizes of the sublattices comprising the topological photonic spin textures. This model not only accounts for photonic spin textures characterized by quasi-periodic structures but also encompasses periodic structures (see Supplementary Section 1), addressing ambiguities in previous research on skyrmion and meron lattices under periodic conditions[28,41]. Among the calculated sublattices of the photonic spin texture, fractal structures can be observed (see Supplementary Section 2), exhibiting a self-similarity property also found in quasi-periodic tiling[42]. Moreover, the model demonstrates distinct characteristics of quasi-periodic photonic spin texture in contrast to periodic circumstance. For example, previous research on periodic photonic spin texture revealed that the spin structure vanished when there was no external angular momentum ($l=0$). However, the spin structure of quasi-periodic photonic spin texture persists even when $l=0$, provided that $N$ is an odd integer, which differs from the phenomenon observed in periodic photonic spin textures (see Supplementary Section 1).

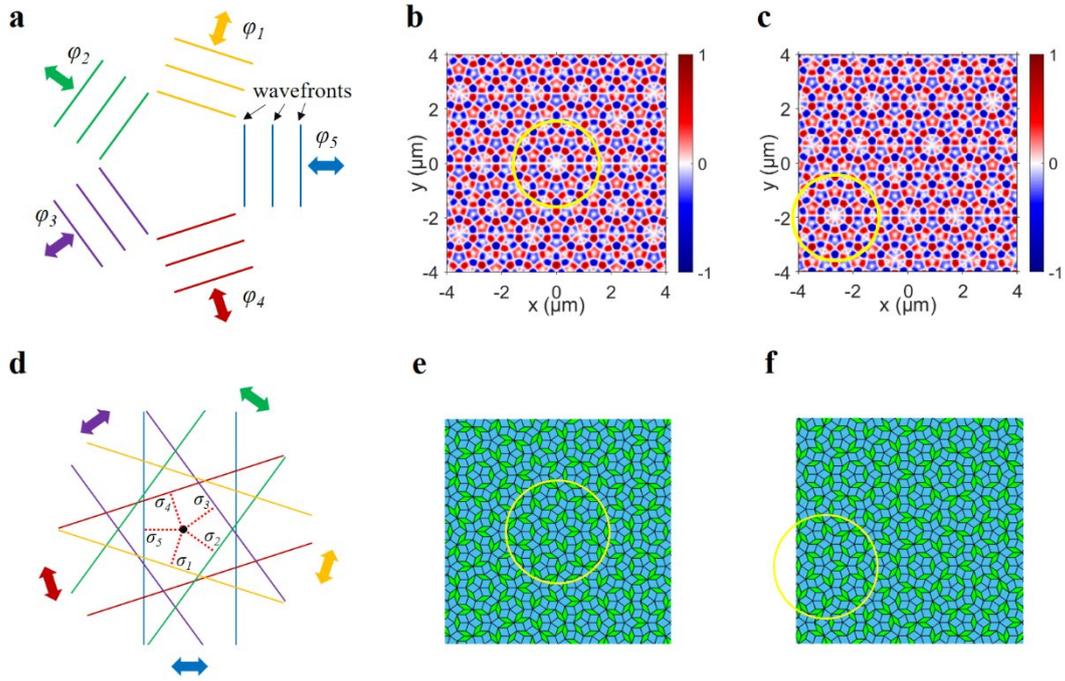

**Fig. 2 | The de Bruijn tiling-style manipulation of the configurations of photonic spin quasicrystals through adjustments of the wavefronts. a,** Schematics of the adjustment of wavefronts of evanescent waves. **b,** The calculated $S_z$ with $N=5$ and the centre marked by the yellow ring. **c,** The calculated $S_z$ with additional phases $(\varphi_1, \varphi_2, \varphi_3, \varphi_4, \varphi_5)=(0, 0, 0.57\pi, 0, 0)$ applied on the evanescent waves. **d,** Schematics of the adjustment of the offsets of parallel lines in the generalized multigrid method. **e,f,** The de Bruijn tiling constructed from the parallel lines with offsets $(\sigma_1, \sigma_2, \sigma_3, \sigma_4, \sigma_5)=(0.4, 0.4, 0.4, 0.4, 0.4)$ (**e**), and $(\sigma_1, \sigma_2, \sigma_3, \sigma_4, \sigma_5)=(0.4, 0.3, 0.2, 0.1, 0)$ (**f**).

The configurations of the photonic spin quasicrystals can be manipulated by adjusting the wavefronts of the evanescent waves, as illustrated in Fig. 2a. The relative positions of the wavefronts can be altered by introducing additional phases to the evanescent waves. For example, the phases $(\varphi_1, \varphi_2, \varphi_3, \varphi_4, \varphi_5)=(0, 0, 0.57\pi, 0, 0)$ are applied to the quasi-periodic photonic spin texture with $N=5$, where $\varphi_n$ is the phase applied on the $n^{th}$ evanescent wave. The manipulation results in a reconfiguration of the photonic spin quasicrystal, which is equivalent to a displacement of the photonic spin texture (Fig. 2b,c). This approach parallels the field of mathematics, where configuration manipulations of the de Bruijn tiling can be achieved by adjusting the offsets of parallel lines, as illustrated in Fig. 2d. The coordinates of the parallel lines can be expressed as $\vec{r}_n \cdot \vec{v}_n = (Z - \sigma_n)d$, where $\vec{r}_n$, $\vec{v}_n = (\cos(2n\pi/N), \sin(2n\pi/N))$, $\sigma_n$ are the coordinate, normal vector, and offset of the $n^{th}$ set of parallel lines respectively; $Z$ is an integer, and $d$ is the line spacing. For example, the de Bruijn tiling constructed from the parallel lines with offsets $(\sigma_1, \sigma_2, \sigma_3, \sigma_4, \sigma_5)=(0.4, 0.4, 0.4, 0.4, 0.4)$ is shown in Fig. 2e. If the offsets are adjusted to $(\sigma_1, \sigma_2, \sigma_3, \sigma_4, \sigma_5)=(0.4, 0.3, 0.2, 0.1, 0)$, the configuration of the de Bruijn tiling is manipulated, demonstrating a displacement feature, as shown in Fig. 2f.

It was also found that if a set of arbitrary phases are applied on the evanescent waves, the quasi-periodic spin structure maintains its structural integrity except for a displacement, which is different from the behaviour observed in periodic spin textures. This phenomenon can be explained through an analysis in the Fourier domain (see Supplementary Section 3). According to this analysis, the displacement of the quasi-periodic spin texture can be derived from the applied phase using the following equation (see Supplementary Section 4):

$$\varphi_n = k_n \sin\left(\frac{n}{N}\pi\right)\Delta x - k_n \cos\left(\frac{n}{N}\pi\right)\Delta y + 2m_n\pi + \varphi_{re}$$
$$\ldots\ldots \quad , \quad (2)$$
$$\varphi_N = \varphi_{re}$$

where $n=1, 2, \ldots N-1$, $m_n$ is an integer, and $\varphi_{re}$ is the reference phase. We first consider $N=5$ as an example. The integer parameters $m_1, m_2, m_3, m_4$ can be determined through numerical method, and the displacements can subsequently be calculated (see Supplementary Section 4). For example, when the phases $(\varphi_1, \varphi_2, \varphi_3, \varphi_4, \varphi_5)=(0.41\pi, 0.57\pi, 1.21\pi, 0.23\pi, 0)$ were applied to the evanescent waves, the displacement of the spin texture were calculated to be $(\Delta x, \Delta y)=(2.359\mu m, -0.3512\mu m)$. The $S_z$ distribution after applying the phases was calculated and is shown in Supplementary Fig. 5, demonstrating consistency with the theoretical prediction. For other odd values of $N$, the displacements of the spin textures can also be calculated from the applied phases by using equation (2) (see Supplementary Section 4). In this context, $N=3$ is an exceptional case, for which the corresponding spin texture is periodic and the displacements can be determined analytically (see Supplementary Section 6).

If the additional phase is applied on only one of the evanescent waves, it can be deduced from equation (2) that the spin texture exhibits one-dimensional displacement along the direction of the corresponding evanescent wave (see Supplementary Section 5). For example, when the phases $(\varphi_1, \varphi_2, \varphi_3, \varphi_4, \varphi_5)=(0, 0, 0, 0, \varphi_{re})$ were applied to the evanescent waves with $N=5$, and $\varphi_{re}=0.5\pi$, $0.57\pi$ or $0.64\pi$, the calculated one-dimensional displacements were $\Delta x=-1.1\mu m$, $3.29\mu m$ and $-3.83\mu m$, respectively. The corresponding results of $S_z$ are shown in Supplementary Fig. 7. It was found that changing the phase of the evanescent wave results in a discontinuous translation of the spin texture, indicating the presence of phason-like dynamics within quasi-periodic systems. Phason excitations are unique to quasi-periodic systems, playing a role similar to phonons in periodic systems. According to the mathematical theory proposed by N. G. de Bruijn, quasi-periodic structures can be viewed as two-dimensional slices of hypercubic structures in higher-dimensions. Consequently, the discontinuous translation of phasons in physical space corresponds to continuous dynamics occurring in higher-dimensional configuration space.

We utilized an in-house developed near-field imaging technique[43] to generate and map the spin structure of the quasi-periodic photonic spin textures. In this experimental scheme, the spin textures of surface plasmon polaritons (SPPs) sustained at the dielectric-metal interface of the sample were characterized using a dielectric nanoparticle as a near-field probe. A SLM was employed to generate $N$ evenly distributed light spots by a $4f$ system which were then focused via an oil-immersion objective lens onto the sample (see the Methods for the details of the experimental set-up). The use of the SLM resulted in significantly smaller light spots on the back focal plane of the objective compared to those produced by intensity masks with apertures in previous studies[28,44]. The incident energy was effectively utilized in this photonic system, and the effective scanning scope was expanded from $2\times2\mu m^2$ to $8\times8\mu m^2$. Furthermore, the phase profile of the SLM can be manipulated by a computer to change the symmetry of the photonic spin textures, which is significantly more convenient than substituting intensity masks with varying apertures.

The measured longitudinal component of SAM for different values of $N$ ($N=5$, $N=7$, and $N=9$) and the phases of $\varphi_n=2\pi ln/N$ with $l=1$ is shown in Fig. 3a–c. The distributions of the measured quasi-periodic spin texture demonstrate $N$-fold rotational symmetry and correspond well to the calculated results for respective symmetries (Fig. 3d–f). The Fourier transform results are shown in the insets of Fig. 3. Except the central point influenced by the background illumination, the experimental results demonstrate $(N-1)/2$ sets and a total of $N^2-N$ points for various $N$ values and are consistent with the theoretical prediction (see Supplementary Section 1).

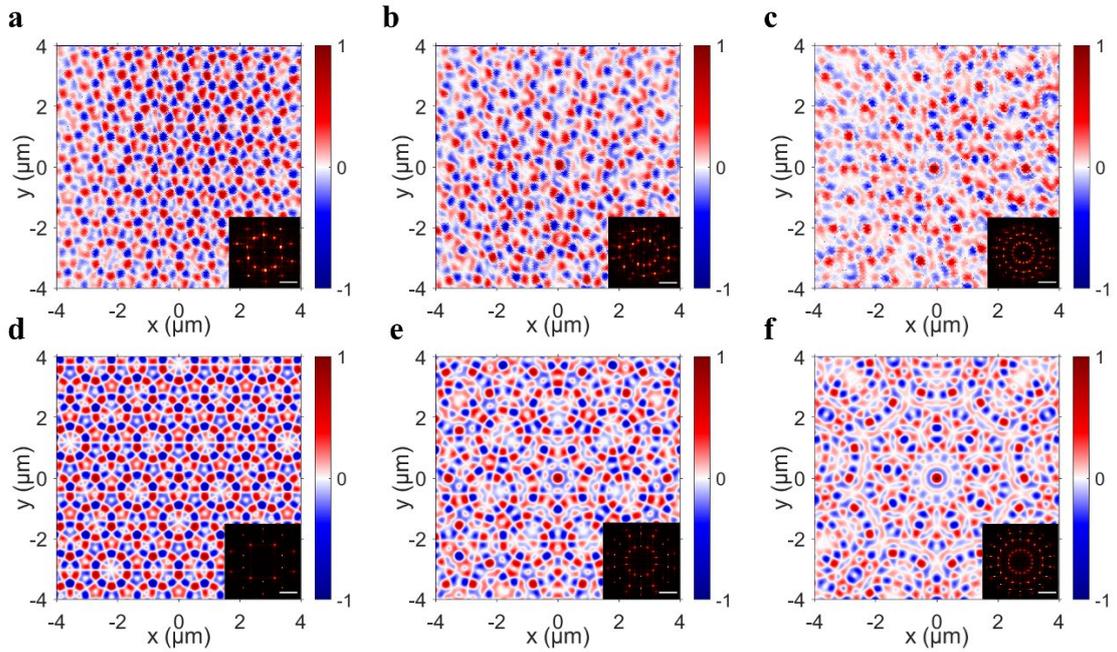

**Fig. 3 | Measurement of photonic spin quasicrystals with diverse symmetries. a–c,** The measured longitudinal component of SAM with $l$=1 and $N$=5 (**a**), $N$=7 (**b**), $N$=9 (**c**). The corresponding Fourier transform results are shown in the insets. **d–f,** The corresponding calculated results of the spin textures. The scale bar in the insets is $k_r$ in the Fourier domain.

Moreover, $S_z$ for $N$=5 and spiral phases with different values of $l$ ($l$=2, $l$=3, $l$=4, and $l$=5) were measured (Fig. 4a–d) and consistent with the calculation results (Fig. 4e–h). In this context, the spiral phase with $l$=5 is equivalent to the situation where no extra phase is present. By manipulating the values of $l$, different phases were applied to adjust the wavefronts of the evanescent waves, while the photonic spin textures maintained structural integrity except for different displacements. These displacements can be obtained from equation (2) and were calculated to be $\Delta x$=0$\mu m$ and $\Delta y$=2.3196$\mu m$, −2.3196$\mu m$, −3.7371$\mu m$, and 0$\mu m$, respectively, for $l$=2, $l$=3, $l$=4, and $l$=5.

It is worth noting that our proposed theoretical framework can be extended to other physical systems. It has been demonstrated that the spin angular momentum in electromagnetic guided waves can be derived from the kinetic momentum as $\boldsymbol{S}=\frac{1}{2k^2}\nabla\times\boldsymbol{\Pi}$, without the need for priori information of the electric and magnetic fields[45,46]. This spin–momentum locking property has shown its universality in various types of fields, such as fluid[47], elastic[48], acoustic[49], and gravitational waves[50], which can be utilized to construct diverse topological spin structures, including Möbius-strip[51] and skyrmions[52–54] in a variety of classical waves. In the future, quasi-periodic spin textures are expected to be constructed based on our theoretical framework in various physical systems. This development is significant for the exploration of wave–matter interactions and the manipulation of the spin degrees of freedom in interdisciplinary research.

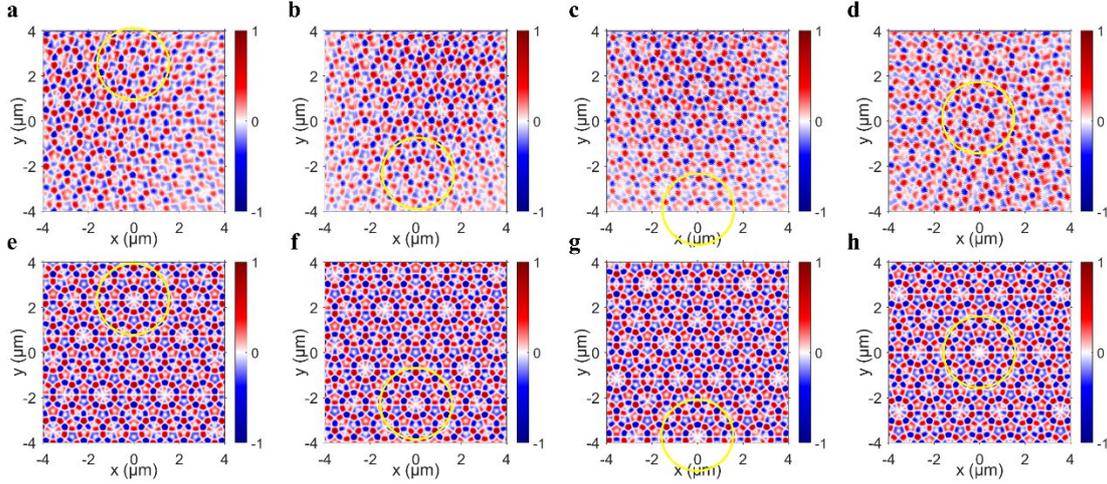

**Fig. 4 | Phason-like discontinuous dynamics with wavefront modulations. a–d**, The measured longitudinal component of SAM with $N=5$ and $l=2$ (**a**), $l=3$ (**b**), $l=4$ (**c**), $l=5$ (**d**). The corresponding Fourier transform results are shown in the insets. **e–h**, The corresponding calculated results of the spin textures.

In summary, we investigated for the first time the quasi-periodic photonic spin textures with diverse rotational symmetries and the novel characteristics arising from spin–orbit coupling under quasi-periodic conditions. We proposed a theoretical framework to elucidated the formation mechanism of photonic spin textures through the interference of evanescent waves. This model aligns with the generalized multigrid method in mathematics, which can be used to generate quasi-periodic tiling with various symmetries. Additionally, the wavefronts of evanescent waves can be adjusted to manipulate the configurations of the quasi-periodic photonic textures by introducing additional phases to the evanescent waves. This method has a mathematical counterpart in the generalized multigrid method, where quasi-periodic tiling is manipulated through the adjustment of parallel line offsets. It was found that changing the additional phase resulted in a discontinuous translation of the spin texture for odd values of $N$, indicating the presence of phason-like dynamics within quasi-periodic systems. Through Fourier domain analysis, the corresponding displacements can be determined theoretically from the applied phases for the first time. This work is expected to be valuable for understanding spin–orbit coupling with various symmetries and configuration manipulations in quasi-periodic photonic systems, with potential applications in optical trapping, quasicrystal fabrications, and optical encryption systems.

## Methods

**Experimental set-up**

In the previous studies of the periodic photonic spin textures of SPPs[28,44], the incident beam was modulated with intensity masks comprised of apertures to break the rotational symmetry and generate the interference of the evanescent waves, which blocked most of the incident energy with only a small portion passing through the apertures. Under such conditions, the angles of the apertures were designed to be larger than 10° to maintain a relative high signal-to-noise ratio of the mapping results, which determined the effective scanning scope of the spin textures to be smaller than $2\times2\mu m^2$ in the previous studies and it is not enough to demonstrate the long-range order of the quasi-periodic photonic spin texture in this work. To expand the scanning scope, a SLM was employed to generate $N$ evenly distributed light spots for interference of the evanescent waves. The experimental set-up is shown in Supplementary Fig. 9. After passing through a telescope system, an incident laser beam with wavelength of 632.8nm illuminated the reflecting liquid SLM, which provided the phase profile of the interference of $N$ waves. The Fourier transform of $e^{i\Phi}$ manifests as $N$ spots which are evenly distributed on a ring. In the experiment, the modulated incident field was weakly focused by a lens with focal length of *500mm* to perform the Fourier transform. The $N$ evenly distributed light spots were generated at the focal point, and then were focused by a *4f* system onto the back focal plane of the oil-immersion objective (Olympus, NA=1.49, 100×). A combination of a linear polarizer (LP) and a *m*=1 vortex wave plate (VWP) was employed to turn the incident field into radial polarized. The incident beam was tightly focused by the oil-immersion objective to excite the SPPs at the air/gold interface of the sample with 50-nm-thick gold film deposited on a glass substrate. By adjusting the wavevectors in the phase profile, the $N$ evenly distributed light spots matched the dark ring which indicated the excitation of the SPPs. In this instance, $N$ evanescent waves were generated at the air/gold interface.

A polystyrene (PS) nanoparticle with a radius of 160 nm was immobilized on the gold film by the 4-mercaptobenzoic acid (4-MBA) molecular linker to scatter the transverse component of the near-field SPPs to the far-field for collection. The scattering radiation was collected by an objective (Olympus, NA=0.7, 60×), and the right and left circular polarized components of the scattering radiation were filtered out by a quarter-wave plate (QWP) and a LP. After polarization selection, the intensity of light of each circular polarized component is measured by the photo-multiplier tube (PMT, Hamamatsu R12829). The nanoparticle-on-film sample was fixed on a Piezo scanning stage (Physik Instrumente, P-545) to perform the two-dimensional scanning, and the spin structure can be mapped.

## Data availability

The data that support the findings of this study are available from the corresponding author on reasonable request.

## Acknowledgements

This work was funded by Guangdong Major Project of Basic Research (Grant No. 2020B0301030009); the National Natural Science Foundation of China (Grant Nos. 62075139, 61935013, 12004260); the Natural Science Foundation of Guangdong Province (Grant Nos. 2024A1515012503, 2023A1515012670); Science and Technology Innovation Commission of Shenzhen (Grant Nos. RCJC20200714114435063, JCYJ20220531103403008); Research Team Cultivation Program of Shenzhen University (Grant No. 2023QNT012); Shenzhen University 2035 Initiative (Grant No. 2023B004).

## Author Contributions

M.L. and L.D. developed the concept of the work. M.L. and Z.X. designed the experiment. M.L., X.G and A.Y. performed the measurements. M.L and L.D. performed the analytical and numerical calculations. M.L. and L.D. wrote the manuscript. L.D. supervised the work. All the authors discussed the results and commented on the manuscript.

## Competing interests

The authors declare no competing interests.